# Slow-light enhanced gain in active photonic crystal waveguides

Sara Ek, Per Lunnemann, Yaohui Chen, Elizaveta Semenova, Kresten Yvind and Jesper Mørk*

DTU Fotonik, Dept. of Photonics Engineering
Technical University of Denmark
DK-2800 Kgs. Lyngby, Denmark
*E-mail: jesm@fotonik.dtu.dk

**Slow light is a fascinating physical effect, raising fundamental questions related to our understanding of light-matter interactions as well as offering new possibilities for photonic devices. From the first demonstrations of slow light propagation in ultra-cold atomic gasses,[1] solid-state Ruby[2] and photonic crystal structures,[3] focus has shifted to applications, with slow light offering the ability to enhance and control light-matter interactions. The demonstration of tuneable delay lines,[4-6] enhanced nonlinearities[7,8] and spontaneous emission,[9] enlarged spectral sensitivity[10] and increased phase shifts illustrate the possibilities enabled by slow light propagation, with microwave photonics emerging as one of the promising applications.[11] Here, we demonstrate that slow light can be used to control and increase the gain coefficient of an *active* semiconductor waveguide. The effect was theoretically predicted[12-14] but not yet experimentally demonstrated. These results show a route towards realizing ultra-compact optical amplifiers for linear and nonlinear applications in integrated photonics and prompts further research into the rich physics of such structures.**

Stimulated emission of radiation leads to the amplification of a propagating electromagnetic wave, forming the basis for optical amplifiers and lasers. The strength of the process is quantified by the gain coefficient per unit length, and is usually considered to be governed by the material properties. Dowling et al.[12] suggested that the slow-down of light near the edge of a 1D photonic bandgap structure could be used to enhance the gain coefficient. The slow propagation of the Bloch mode near the band edge, which may be visualized as multiple back-and-forth scattering of the light beam, thus lengthens the local dwell time in the medium and increases the spatial but not the temporal gain coefficient.[15] Despite obvious applications in optical amplifiers[16] and lasers,[13] there is no conclusive experimental demonstration in a waveguide structure. For dye-doped 3D artificial opals, enhancement of gain in certain crystallographic directions and its relation to the directional density of optical states was demonstrated[17] and in Ref. 18 enhancement effects in multimode InP waveguides related to a ministopband were indicated. Spontaneous emission enhancement was observed in photonic crystal wires[19] and dielectric multilayers with oxygen vacancies as light emitters[20] and may be used to increase the spontaneous emission factor of a waveguide.[9] Early measurements on lasers employing photonic crystal waveguides[21] and coupled resonator structures[22] show that the lasing characteristics are improved using slow light effects. However, since both the cavity quality factor and the spatial gain coefficient may be affected by slow light propagation, laser measurements cannot easily distinguish between the two effects. Thus, measurements on open structures are required.

Before describing our experimental results, we briefly summarize the theoretical status. An extensive analysis for a periodic photonic crystal lattice shows that the gain scales with the density of optical states.[13] For a one-dimensional defect waveguide, based on qualitative considerations of the effective path length, Dowling et al.[12] suggested a linear scaling of gain with group index. By expansion into Bloch waves we derived the following approximate expression for the modal gain coefficient (see supplementary information)

$$g_{\mathrm{mod}} = \Gamma g_{\mathrm{mat}} = \Gamma \frac{n_g}{n_b} g_0 \qquad (1)$$

with $g_{\mathrm{mat}}$ being the material gain, $g_0$ the bare material gain in absence of a PCW, $n_g$ the group index, $n_b$ the background refractive index of the active material, and $\Gamma$ a generalized confinement factor describing the field overlap with the active material. The importance of the latter was demonstrated experimentally in absorption-type measurements.[23] Equation (1) agrees with the result of a perturbative analysis.[24] The predicted scaling of the modal gain (or absorption) with group index has, despite its intuitive appeal, been the subject of some controversy, and it was shown that one needs to distinguish between slow light originating from structural perturbations, e.g. a photonic crystal (PhC) waveguide, and material dispersion, such as electromagnetically induced transparency.[25] For the latter, equation (1) does not hold, as also demonstrated experimentally.[26] Also, for translationally invariant systems, such as a fibre, the gain enhancement due to slow-light induced by mode dispersion is exactly cancelled by the resulting transversal expansion of the mode.[25]

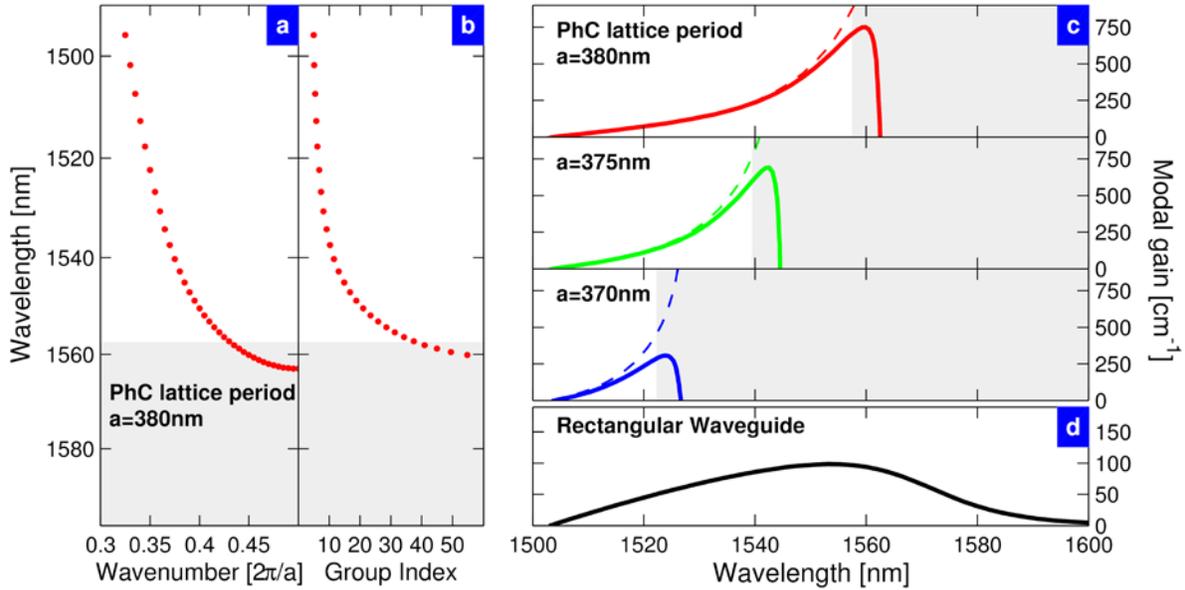

**Figure 1| Comparison of calculated modal gain in standard and slow-light enhanced waveguides. a,** Photonic band diagram and **b**, corresponding group index of the passive 3D PhC waveguide for TE-like modes. (Triangular lattice of air holes with a=380 nm, r=0.26a, h=340 nm, $n_b^2$=11.2, ten layers of 10 nm QWs placed in the middle of membrane). In the grey area the gain is impacted by slow-light enhanced scattering loss due to disorder. **c**, Modal gain (dashed curves: neglecting disorder) and net gain (solid curves: including disorder-induced losses) of fundamental TE-like modes for three 10 QW PhC waveguides with lattice period of 380nm (upper), 375nm (middle) and 370nm (lower row). **d**, Modal gain of fundamental TE-like mode for a reference rectangular waveguide with ten QW layers (width 660nm, height 340nm, $n_b^2$=11.2, group index 3.8, and confinement factor of 40%).

## Results

To confirm (1) and quantify the impact of slow light on the modal gain we carried out extensive calculations of gain spectra in different structures, see Fig. 1. As the mode frequency (Fig. 1a) approaches the band-edge, the group index (Fig. 1b) of the PhC waveguide strongly increases, leading to a significant increase of the modal gain coefficient of the corresponding active PhC waveguide (Fig.1c, a=380 nm). Gain spectra for two other PhC lattice constants and the same electronic density of states and carrier distributions are also shown. The gain spectra may be compared to that (Fig. 1d) of a rectangular waveguide (nanowire) embedded in air, which is proportional to the material gain spectrum and has a confinement factor as high (approximately 40 %) as the PhC structures (see supplementary information for additional detail). It is well-known that as the band-edge is approached, scattering losses are enhanced due to slow-light propagation.[27-29] This loss enhancement was included phenomenologically, following Ref. 28, leading to disorder dominated losses in the regions shaded grey in Fig. 1c. Through-out the remainder of this work, we shall refer to the modal gain minus waveguide losses as the net gain, $g$. These results thus predict that the net gain can be enhanced and controlled via the photonic environment rather than the usual approach of modifying the epitaxial structure.

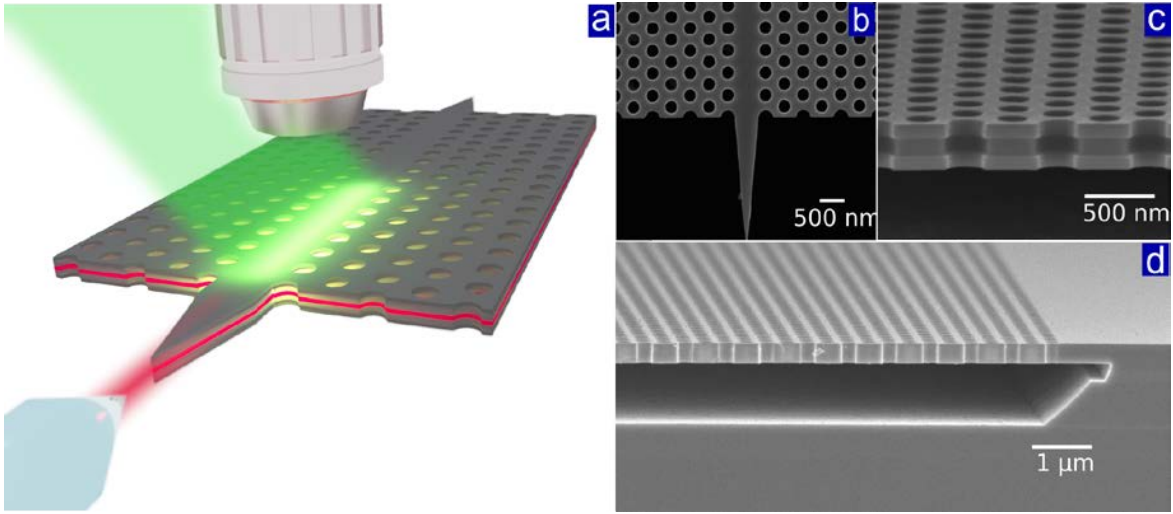

**Figure 2 | Measurement set-up and structures investigated.  a,** Schematic of the active membrane exposed to a focused pump beam (green) and the fibre collecting light from the taper. SEM images showing **b**, the taper, **c**, the active material in the centre of the membrane and **d**, the PhC membrane.

PhC waveguides with one missing air hole (W1-type) are fabricated in a 340-nm thick air-embedded InGaAsP membrane with a triangular-lattice pattern of air-holes, see Methods. The waveguide is terminated by tapers[30] to suppress residual reflections and improve coupling. Similar *passive* structures have been used for investigating slow-light propagation[3,4] and nonlinearities[7,8] but in our case an active material is incorporated inside the membrane and the structures are designed such that the slow-light regime spectrally overlaps with the region where positive net material gain can be achieved by optical pumping. A schematic of the setup as well as SEM pictures of fabricated samples are shown in Fig. 2. Fig. 3 shows measured spontaneous emission spectra for three 1 mm long waveguides with different PhC lattice parameters, corresponding to a progressive wavelength shift of the slow light region by 20 nm.

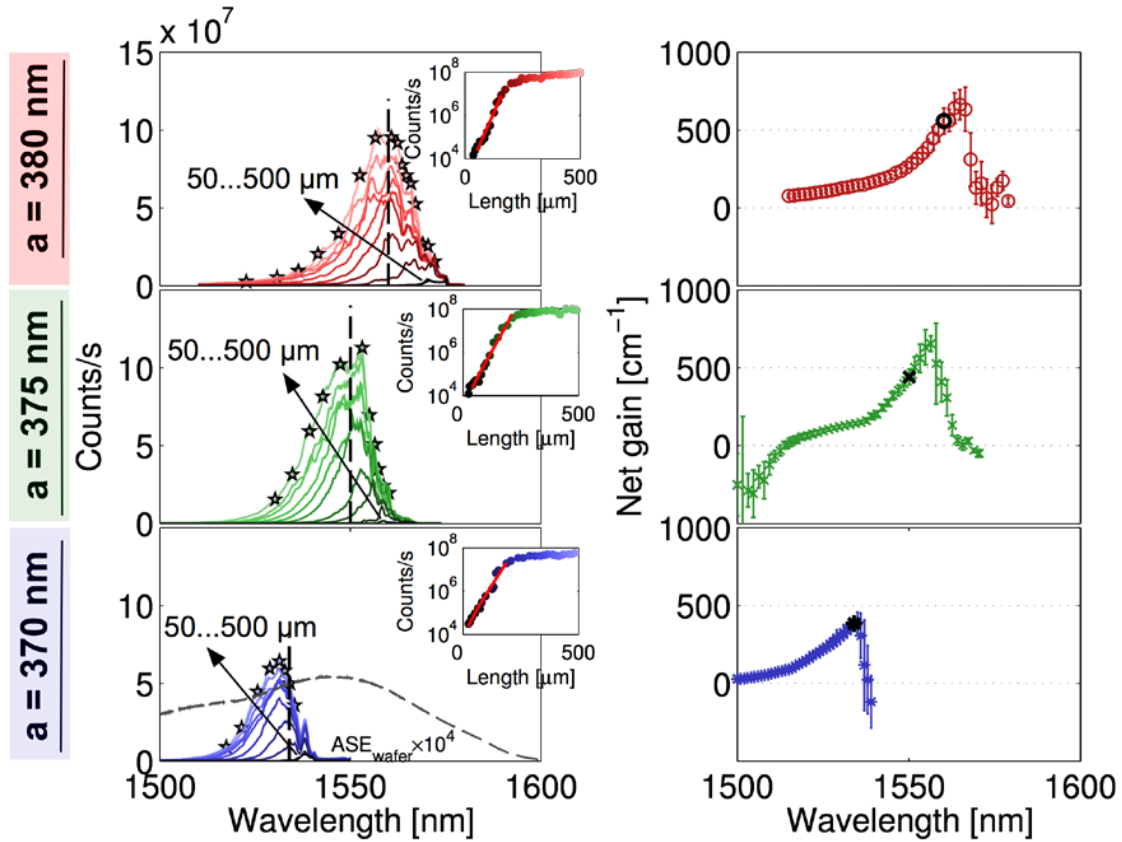

**Figure 3 | Slow-light enhanced spontaneous emission and gain spectra.** Amplified spontaneous emission spectra (left column) for varying pump stripe length (steps of 50 µm), and derived net gain vs. wavelength (right column) for three 10 QW PhC waveguides with lattice constants of 380 nm (upper), 375 nm (middle) and 370 nm (lower row). The insets show the intensity versus stripe length and the corresponding fits for the wavelengths indicated by vertical dashes lines. The extracted gain value is indicated by a black marker in the gain spectrum (right). The star-markers (left column) indicate spectral features in the spectra that are used for extracting an optical path length (cf. Fig. 4).

The devices have identical 10 QW epitaxial structures and the measurements thus clearly show that the emission spectrum can be controlled via the PhC lattice parameters. The different spectra shown for each of the samples correspond to different excitation stripe lengths, defined by blocking part of the pump light in close vicinity to the sample. If the modal gain exceeds the total losses, the spontaneous emission is amplified along the waveguide. To measure the net gain, including propagation losses due to non-resonant residual absorption as well as disorder-induced scattering, we use a standard method originally introduced in the seminal work of K. L. Shaklee *et al.*[31] by fitting the data to the expression

$$I(l) = \frac{A}{g}(e^{gl} - 1), \tag{2}$$

where the output intensity, $I$, depends on the excitation stripe length, $l$, and $A$ is proportional to the local rate of spontaneous emission. Inserts in Fig. 3 show example variations of intensity with stripe length, with the intensity increasing by more than three orders of magnitude, and the corresponding fits. For long stripe lengths the intensity saturates, presumably due to saturation of the gain as for standard semiconductor optical amplifiers, and this part is omitted in the fit (see Methods for details). Fig. 3 shows the extracted gain spectra for the three waveguides under similar pumping conditions. The spectra are seen to shift in good qualitative agreement with the simulations shown in Fig. 1c. Our measurement scheme avoids the problem of Fabry-Perot resonances induced by facet reflections, since the far-side facet is preceded by un-pumped material with strong absorption. The caveat is that due to the absence of a well-defined cavity, we do not have access to independent measurements of the absolute group index variation of the structures.

To confirm that our waveguides do display the generally accepted variation of group index with wavelength[27-29], we present in Fig. 4 the measured emission spectrum of a structure with a=380 nm and r/a=0.25 (same as the waveguide plotted in red in Fig. 3) but implementing a 10 μm long PhC cavity by terminating the waveguide at both ends by holes. A ten-fold light slow-down is observed close to the band edge of this device. The slight shift in band edge position, compared to the samples in Fig. 3, is related to the lower relative permittivity of this waveguide, which contains one layer of quantum dots instead of 10 quantum wells. To get information on the location of the band edge for the exact same samples as investigated in Fig. 3, we analyze the fringe spacing, $\delta\lambda$, in the spectra, most clearly noticed for the sample with a=380 nm close to the band edge. We attribute these fringes to imperfections in the lattice that cause reflections. Because the length of the waveguide section causing the modulation is unknown, we only extract the corresponding variation of the path length, *$n_g L = \lambda^2/(2\delta\lambda)$*, see Fig. 4 (top). The position of the band edge extracted from these measurements correlate well with the position at which the measured gain spectra attain their maxima, cf. Fig. 3, thus confirming that the gain enhancement is a slow-light effect.

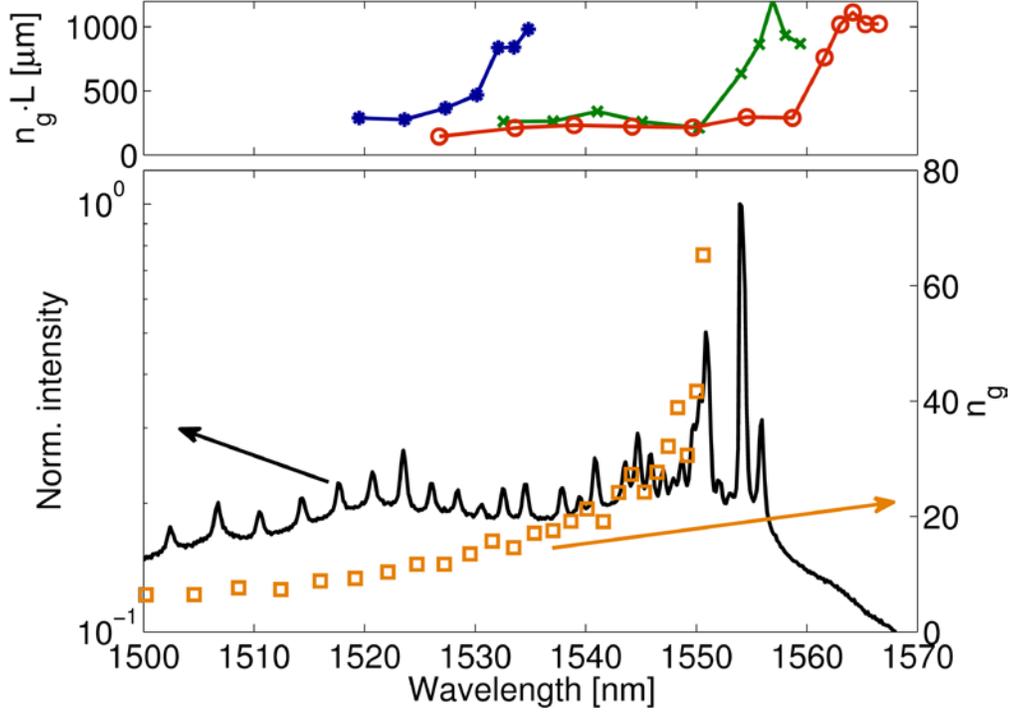

**Figure 4| Measured group index and position of band-gap edge.** Bottom: Measured amplified spontaneous emission spectrum and group index for a 10 µm long cavity in a PhC slab with a=380 nm and r/a=0.25. Top: variation of path lengths extracted from emission spectra shown in Fig. 3 for the lattice constants 380 nm (red circles), 375 nm (green, crosses), and 370 nm (blue asterisks).

## Discussion

With present technology, passive PhC waveguides exhibit a propagation loss of about 2 dB mm$^{-1}$ in the conventional "fast-light" region and follows approximately an $n_g^2$ scaling.[28] Even though the propagation loss therefore is much higher than for well-designed ridge waveguides, a positive net gain coefficient up to about 700 cm$^{-1}$ is measured (Fig. 3). Standard ridge type waveguides, with 1 QW of the same composition and thickness as employed here, were found to have a maximum modal gain of ~15 cm$^{-1}$, and we therefore observe a total gain enhancement factor of order 700/(10×15)=4.7, clearly larger than the factor ~2 increase expected from the tighter confinement of the waveguide (see supplementary information). We believe that in the present device, besides the propagation losses mentioned above, the net gain is limited by heating and carrier loss due to surface recombination. However, recent work demonstrates the feasibility of realizing a buried heterostructure and electrical pumping,[32] which will strongly reduce these limitations. Furthermore, dispersion engineering[4] may be applied to optimize the group index variation relative to the gain peak.

From the gain spectra in Fig. 3 one notices that the transparency point (where the net gain is zero) appears blue-shifted for the smallest PhC lattice periodicity. This indicates that the resulting carrier density is higher for that structure despite the near-identical pumping conditions and is likely due to suppression of spontaneous emission at longer wavelengths caused by the photonic bandgap. This suggests a way of increasing the quantum injection efficiency into the desired levels without changing the electronic level structure and would be very interesting from a pump efficiency point of view. So-far, such tailoring was accomplished by exploiting quantum confinement effects in low-dimensional semiconductors. Further investigations with improved measurement accuracy and implementation of electrical pumping are, however, required to conclude on this.

Similar to conventional semiconductor optical amplifiers, the gain of a PhC amplifier will be limited by carrier depletion caused by the injected signal or by amplified spontaneous emission. Slow-light propagation may cause additional effects. As already mentioned, random scattering due to fabrication-induced disorder reduces the maximum attainable group index and increases the loss. As the band-edge is approached, local cavities may even form, leading to random lasing.[33] Even in the absence of disorder, distributed feedback (band-edge) lasing[21] may set in close to the Brillouin zone edge and these lasing effects will lead to further carrier depletion thus limiting the maximum gain. It was recently predicted that the slow-down of the light itself will be limited by the presence of gain, due to broadening of the optical density of states.[34] This effect, not yet experimentally observed, may also be contributing to the gain saturation that we observe experimentally, but we cannot presently separate that from the effect of gain depletion. Practical applications will require electrical pumping and efficient heat removal, but recent progress on photonic crystal membrane lasers is very promising in this respect.[32]

In conclusion, we have demonstrated that the gain of a waveguide can be enhanced and spectrally controlled by exploiting slow light propagation. This is promising for the realization of compact integrated photonic crystal amplifiers, which are essential, presently missing, components for integrated photonic chips. The use of slow light may also reduce the power consumption of amplifiers since less active material needs to be inverted to achieve a given gain. The control of the gain profile via the photonic structure, rather than the epitaxial structure, furthermore provides new design freedom, e.g. allowing varying the gain coefficient along the waveguide, which may be exploited for ultrafast low-power signal processing applications. The present work also demonstrates a platform for fundamental studies of propagation and random scattering effects in structures under the combined effect of slow light and net gain; topics, which have proved very rich, even for passive structures.

## Methods

**Device fabrication.** The wafers were grown by metal-organic vapour phase epitaxy on InP substrates. The membrane is InGaAsP ($\lambda_g$=1.15 µm) with a total thickness of 340 nm containing one or ten strain compensated quantum wells (6nm $In_{0.75}Ga_{0.25}As_{0.86}P_{0.14}$ well, 7.2nm $In_{0.49}Ga_{0.51}As_{0.86}P_{0.14}$ barrier) in the center. Sacrificial layers underneath were formed by a stack of 100 nm InP, 100 nm InAlAs and 800 nm InP with a total thickness of 1 µm. A 200 nm layer of $SiN_x$ was deposited by PECVD followed by a 500 nm thick layer of a positive e-beam resist (zep520A). The patterning was done using e-beam writing by a JEOL-JBX9300FS. The pattern was transferred to the $SiN_x$ by CHF3/O2 RIE and further transferred to the semiconductor by cyclic $CH_4/H_2 - O_2$ RIE after resist removal. Beneath the active slab there is a 1 µm thick sacrificial layer, which is etched away using $1HCl:2H_2O$ selective wet-etching in order to obtain an air-slab structure.

**Device characterization.** The devices are optically excited along a stripe by focusing pump laser light onto the waveguide using a cylindrical lens. In order to avoid effects of heating, the samples are pumped with pulses from a mode-locked Ti:Sapphire laser (wavelength 800 nm, duration ~1 ps, repetition rate 270 kHz) with an average power of 6 mW (corresponding to 40 W/cm² average power or a pulse fluence of ~1.5 pJ/µm²). The stripe width was approximately 15 µm. The waveguide output is coupled to a lensed fibre using a PhC waveguide taper. A series of spontaneous emission spectra from well defined excitation stripe lengths are acquired using a cooled InGaAs array detector on a spectrograph. The excitation stripe lengths are measured from calibrated camera microscope images of the spontaneously emitted near infra red (NIR) light scattered out from the top of the waveguide.

**Gain extraction and error bar estimation.** The gain spectrum was extracted by fitting to eq. (2) using the maximum likelihood method. Shot noise was found to be negligible and the noise rather resulted from fluctuations due to waveguide-fiber coupling. Accordingly, since the statistical error is proportional to the mean output signal, we assume a lognormal (rather than a normal) distribution corresponding to the logarithm of the output signal being normally distributed. The fits thus minimize the sum of squared residuals for the logarithm of the measured ASE. The error bars shown are the calculated 95% confidence intervals. It might be expected that the uncertainty should be at a minimum where the ASE spectrum peaks, since here the output signal shows the largest variation with stripe length. However, the signal is observed to saturate at a shorter stripe length for higher gain, limiting the useful fitting range where the exponential dependence in equation (2), which neglects saturation, holds true. Examples of fits with indication of the fitting range employed are shown in the insets of figure 3. In order to systematically determine the fitting range, we carried out fits for all wavelengths versus maximum pump lengths and used the set of wavelengths and fitting ranges resulting in an $R^2$ value (coefficient of determination), closest to 1. $R^2$ is a standard measure for the goodness of a fit, and was evaluated as $R^2 = 1 - S_{err}/S_{tot}$, where $S_{err} = \sum_i [\ln(y_i) - \ln(I_i)]^2$ and $S_{tot} = \sum_i [\ln(y_i) - \langle \ln(y) \rangle]^2$ and $y_i(I_i)$ is the measured ASE with pump stripe length $z_i$.

## Acknowledgment

We thank M. Schubert, University of Konstanz, for developing the photonic crystal membrane process, W. Xue, Technical University of Denmark, for performing the measurement of the spectrum reported in Fig. 4, and A. de Rossi and S. Combrié for help with the taper technology. Villum Fonden is acknowledged for financial support via the NATEC Centre of Excellence.

## Authors contributions

S.E., E.S. and K.Y. fabricated the samples. S.E. and P.L. carried out the optical experiments. Y.C. and J.M. developed the theory. J.M. S.E., Y.C., and P.L. wrote the manuscript. All authors read and commented on the manuscript. K.Y. and J.M. supervised the project.

## Competing financial interests

The authors declare no competing financial interests.

# Supplementary Information

## Theory & Numerical Simulations

Using the Lorentz reciprocity theorem and incorporating the material gain as a small perturbation in finite sections over a passive periodic PhC waveguide background, we have derived a set of coupled equations for the slowly varying envelopes of forward and backward propagating Bloch waves,[1] in analogy with the analysis of Sipe[2] for a periodic Kerr medium based on $\mathbf{k} \cdot \mathbf{p}$ perturbation theory. For sufficiently small material gain, the two equations de-couple and the effective gain for the amplitude of the forward propagating Bloch wave can be expressed in the simple form

$$g_{\mathrm{mod}} = \Gamma g_{\mathrm{mat}} = \Gamma \frac{n_g}{n_b} g_0 \tag{S1}$$

with $g_{mod}$ being the modal gain of the active PhC waveguide, $g_{mat}$ the material gain, $g_0$ the bare material gain in absence of a PCW, $n_g$ the group index in the propagation direction, and $n_b$ the background refractive index of the active material. Furthermore, $\Gamma$ is a generalized confinement factor for the stored electric energy inside the active region of the waveguide:

$$\Gamma = \frac{2 n_b^2 \int_V dV \varepsilon_0 F(r) |\mathbf{e}(r,\omega)|^2}{\int_V dV \left[ \varepsilon_0 \chi_b(r) |\mathbf{e}(r,\omega)|^2 + \mu_0 |\mathbf{h}(r,\omega)|^2 \right]} \equiv \frac{n_b^2 \int_V dV \varepsilon_0 F(r) |\mathbf{e}(r,\omega)|^2}{\int_V dV \varepsilon_0 \chi_b(r) |\mathbf{e}(r,\omega)|^2} \tag{S2}$$

Here, $\mathbf{e}(r, \omega)$, $\mathbf{h}(r,\omega)$ are the normalized electric and magnetic fields of the periodic guided Bloch mode of the passive structure at frequency $\omega$ with propagation constant $\beta$. Also, $F(r)$ is the active material distribution function, $\varepsilon_0$ and $\mu_0$ are the permittivity and permeability of free space, $\chi_b(r)$ is the background susceptibility and $V$ is the volume of a supercell. The general result expressed by Eqs. (S1) and (S2) reduces to the well-known expression for the modal gain of conventional longitudinally invariant ridge waveguides, with the confinement factor, (S2), reducing to an integral over the transverse cross section.

In order to quantitatively compare the modal gain of PhC waveguides with conventional ridge waveguides, we have used the finite element method (FEM) based solver COMSOL to calculate the propagating modes in both passive and active waveguide structures, i.e., 2D modal analysis for ridge waveguides and 3D vectorial-field supercell simulation for PhC waveguides.[3]

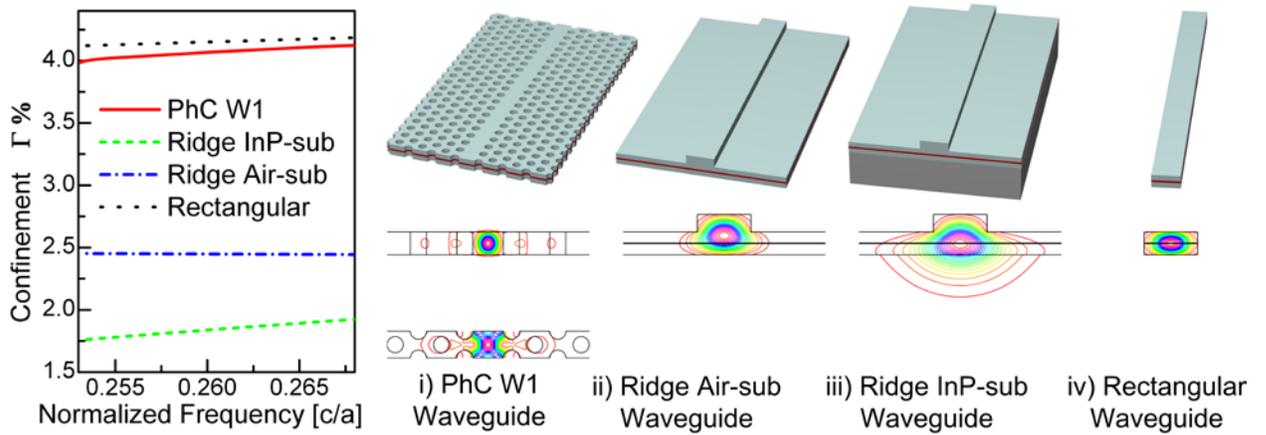

**Figure S1 Confinement factor $\Gamma$ of fundamental TE-like guiding modes in W1-defect InGaAsP PhC membrane waveguide and three reference conventional waveguides.** Four waveguide structures are compared: i) PhC W1 membrane waveguide; ii) (Ridge Air-sub) Ridge waveguide with air cladding on both top and bottom; iii) (Ridge InP-sub) Ridge waveguide with InP substrate; iv) (Rectangular) Rectangular waveguide with air cladding. Side- and top-views of time-averaged electric energy density profiles at $\lambda = 1550 nm$ are shown. Parameters: a=398nm, r=0.3a, h=340nm, $n_b^2$=11.2, 10 nm single QW layer is the middle of membrane. The widths of the conventional waveguides are 2a.).

Fig. S1 shows the calculated confinement factor of the fundamental TE-like guided modes in a PhC W1 waveguide and three conventional waveguides. Examples of typical time-averaged electric energy density profiles for the different structures are shown. The relevant guided modes in the PhC membrane waveguide are tightly confined horizontally by the band gap and vertically by index guiding. The PhC membrane waveguide has a confinement factor similar to the simple rectangular waveguide with air cladding and shows higher confinement factor compared to conventional ridge waveguides employing a semiconductor substrate or even air below the guiding layer.

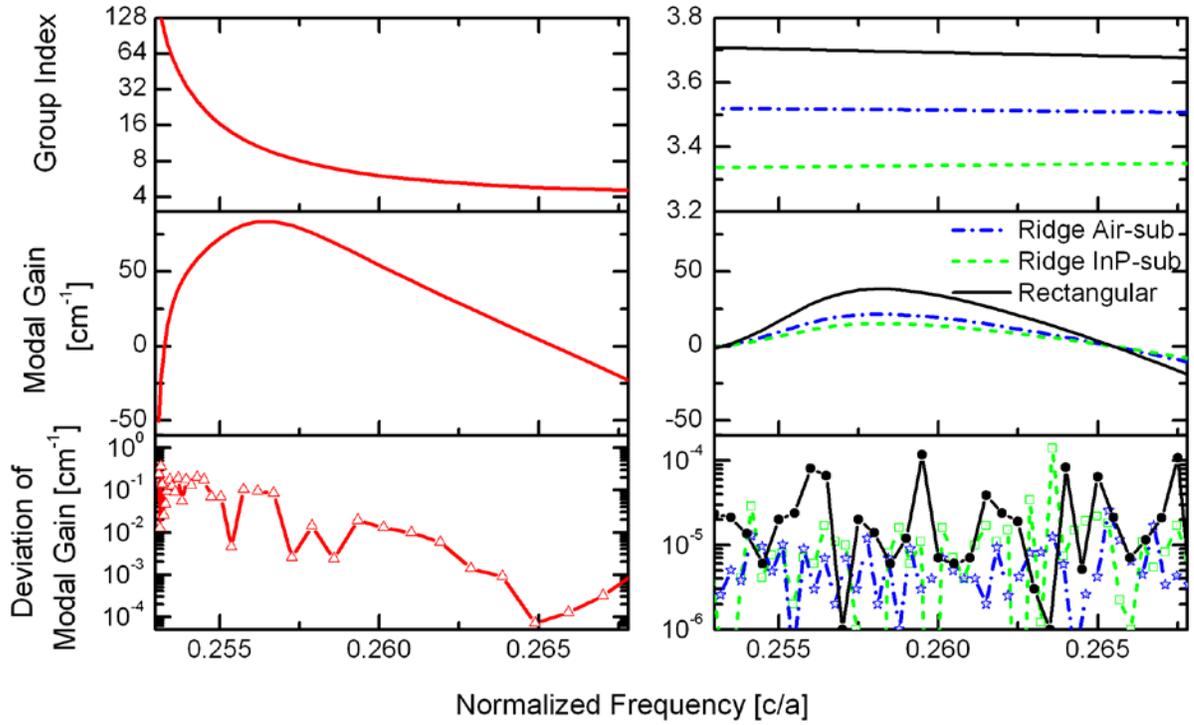

**Figure S2 Modal gain and group index for the four structures.** The frequency dependence of group index, modal gain and the deviation between the analytical expression and the full numerical solution are shown for the PhC waveguide (left column) and the three conventional waveguides (right column). Parameters: a=398nm, r=0.3a, h=340nm, $n_b^2$=11.2, 10nm single QW layer is the middle of membrane. Widths of conventional waveguides are 2a.

Fig. S2 compares the frequency dependence of the group index and modal gain for the same four structures. It shows that while the tighter confinement of the transverse mode in the PhC waveguide contributes to the gain enhancement, it is dominated by the slow-light effect via the variation of the group index. As shown in the bottom row of Fig. S2, the agreement between the perturbative expression for the modal gain, Eq. (1), and the imaginary part of the (numerically exact) wavenumber obtained from FEM simulations of the full active waveguides is excellent. The deviation is smaller than $\sim 10^{-4}$ cm$^{-1}$ for the conventional waveguides and on the order of or smaller than $10^{-1}$ cm$^{-1}$ for the PhC waveguide. The larger error observed for the PhC waveguide is attributed to the larger discretization error encountered for the 3D vectorial-field solutions as well as limitations to the super-cell approximation as the cut-off frequency is approached.

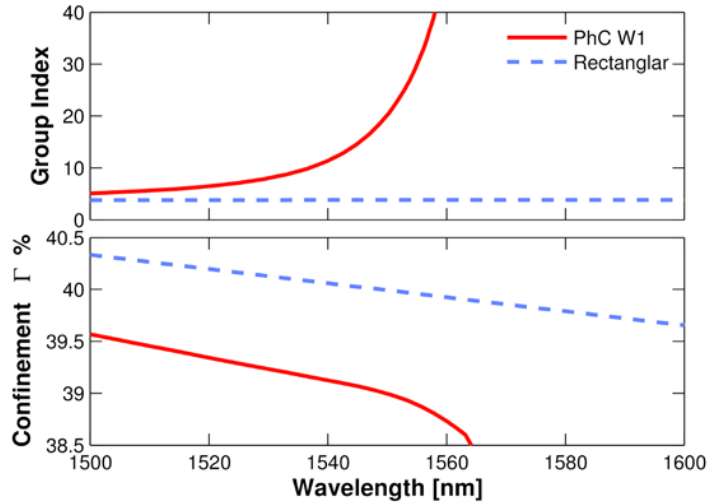

**Figure S3 Comparison of group index and confinement factor for PhC waveguide and rectangular waveguide.** Triangular lattice of air holes with a=380 nm, r=0.26a, h=340 nm, $n_b^2$=11.2, 100nm active layer (10 QWs) place in the middle of membrane. Air-cladding rectangular waveguide with width of 660 nm and height of 340 nm, $n_b^2$=11.2.

Fig. S3 compares the group index and confinement factor for the fundamental TE-mode obtained for the PhC W1 waveguide and the air-cladding rectangular waveguide. Both waveguides have 10 QW layers in the middle and show similar confinement factors of around 40%. For simplicity, in the main text, we use this rectangular waveguide as a reference waveguide, implying a conservative estimate of the enhancement factor attainable.

### Transmission Measurements

Transmission measurements were also carried out. In order to reduce the device length to 200 μm, while avoiding the technological difficulty of cleaving very short devices, the waveguides were terminated with a PhC mirror[4] from which point the waveguide is filled with holes. Only one taper is used for in- and out-coupling using a fiber circulator. The full length of the waveguides is pumped as described in Methods. Using lock-in detection, the pump-induced change of transmission of an injected 1 mW CW beam was measured for a range of wavelengths. Since it is the *change* of transmission that is detected for a fixed length of the waveguide, this method does not allow measuring the absolute gain, as for the measurements presented in Fig. 1. Fig. 4 shows measurement results for waveguides with band edges located at 1600, 1588, and 1566nm. The maximum of the pump-induced transmission change is observed to correlate very well with the slow-light region of the waveguide, confirming the conclusions from the measurements of amplified spontaneous emission that the gain spectrum can be controlled by the photonic crystal parameters.

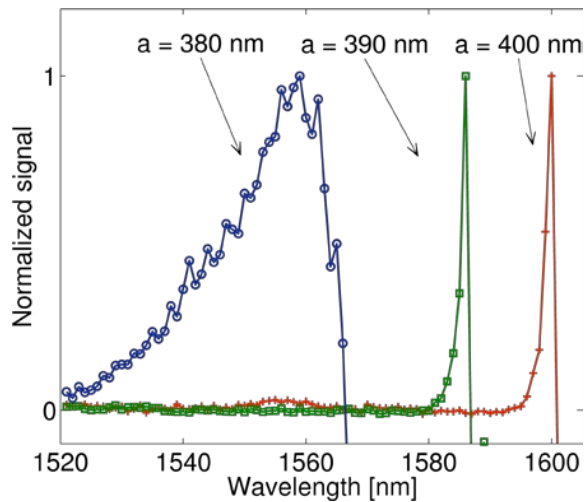

**Figure S4  Enhanced transmission in slow-light region.** The pump-induced change of transmission is shown as a function of the wavelength of the injected signal for three structures with different hole periodicity, a. The wavelength region of high transmission correlates with the slow light region of each crystal. The signals are normalized by their peak value.

### References (Supplementary Information)